\documentclass[twocolumn]{aastex631}
\usepackage{amssymb}
\usepackage{array}
\usepackage{multirow}
\usepackage{bold-extra}

\graphicspath{{./}{figures/}}
\AtBeginDocument{
}

\newcommand{\n}{\nodata}

\received{2022 Oct 3}
\revised{2022 Oct 28}
\accepted{2022 Nov 3}

\submitjournal{ApJ Letters}

\shorttitle{Galactic high-energy neutrinos}
\shortauthors{Kovalev, Plavin, Troitsky}

\begin{document}

\title{Galactic contribution to the high-energy neutrino flux found in track-like IceCube events}

\correspondingauthor{S.~V.~Troitsky}
\email{st@ms2.inr.ac.ru}

\author[0000-0001-9303-3263]{Y.~Y.~Kovalev}
\affiliation{Max-Planck-Institut f\"ur Radioastronomie, Auf dem H\"ugel 69, 53121 Bonn, Germany}
\affiliation{Astro Space Center of Lebedev Physical Institute, Profsoyuznaya 84/32, 117997 Moscow, Russia}
\affiliation{Moscow Institute of Physics and Technology, Institutsky per. 9, Dolgoprudny 141700, Russia}

\author[0000-0003-2914-8554]{A.~V.~Plavin}
\affiliation{Astro Space Center of Lebedev Physical Institute, Profsoyuznaya 84/32, 117997 Moscow, Russia}

\author[0000-0001-6917-6600]{S.~V.~Troitsky}
\affiliation{Institute for Nuclear Research of the Russian Academy of Sciences, 60th October Anniversary Prospect 7a, Moscow 117312, Russia}


\begin{abstract}
Astrophysical sources of neutrinos detected by large-scale neutrino telescopes remain uncertain. While there exist statistically significant observational indications that a part of the neutrino flux is produced by blazars, numerous theoretical studies suggest also the presence of potential Galactic point sources. Some of them have been observed in gamma rays above 100~TeV. Moreover, cosmic-ray interactions in the Galactic disk guarantee a diffuse neutrino flux. However, these Galactic neutrinos have not been unambiguously detected so far.
Here we examine whether such a Galactic component is present among the observed neutrinos of the highest energies.
We analyze public track-like IceCube events with estimated neutrino energies above 200~TeV. We examine the distribution of arrival directions of these neutrinos in the Galactic latitude $b$ with the help of a simple unbinned, non-parametric test statistics, the median $|b|$ over the sample.
This distribution deviates from that implied by the null hypothesis of the neutrino flux isotropy, and is shifted towards lower $|b|$ with the p-value of $4\times 10^{-5}$, corresponding to the statistical significance of $4.1\sigma$.
There exists a significant component of the high-energy neutrino flux of Galactic origin, matching well the multi-messenger expectations from Tibet-AS$\gamma$ observations of diffuse Galactic gamma rays at hundreds of TeV. Together with the previously established extragalactic associations, the Galactic component we report here implies that the neutrino sky is rich and is composed of contributions from various classes of sources.
\end{abstract}


\section{Introduction}
\label{sec:intro}

High-energy (few TeV to few PeV) astrophysical neutrinos were discovered by IceCube \citep{IceCube-2013} and now confirmed by all three instruments capable of their detection \citep{IceCube-HESE-2020,ANTARES-diffuse,Baikal-Neutrino2022a}. Their origins are not easy to be determined, mainly because of the huge background of atmospheric events and because of the poor angular resolution of the neutrino detectors. While numerous models of astrophysical neutrino emitters were proposed and discussed, they generally fail to fit the entire ensemble of observational data, including neutrino spectra, distribution of arrival directions and multi-messenger constraints, in terms of a single population of sources \citep[see e.g.][for a recent review]{ST-UFN}. In particular, under the assumption of the extragalactic origin of the entire neutrino flux, some tension arises between IceCube cascade neutrino spectra and \textit{Fermi}~LAT \citep{FermiDiffuse} diffuse gamma-ray fluxes \citep[see e.g.][and references therein]{IceCube-HESE-2020}. This long-standing tension lead to the suggestion that the total neutrino flux is not saturated by a single class of sources but includes both Galactic and extragalactic contributions \citep{2comp-Chen,2comp-Vissani,2comp-Neronov,2comp-Vissani-2}. 
Population studies provide growing statistical evidence that the extragalactic component is associated with parsec-scale emission in blazars \citep{neutradio1,neutradio2}, see also \citet{Resconi2020,Hovatta2021,
Aublin2021,Illuminati2021,
Franckowiak2022,Buson}.

\begin{figure*}
\centering
\includegraphics[width=0.7\linewidth]{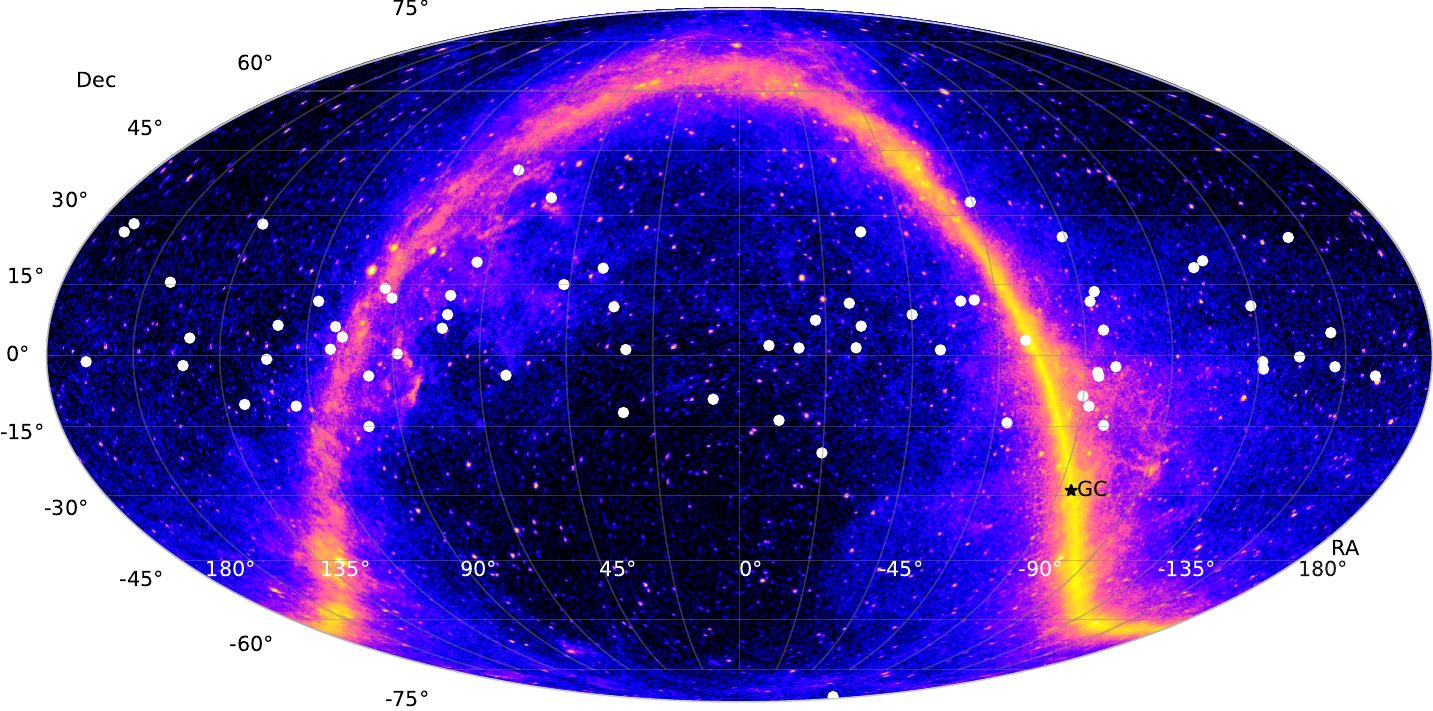}
\caption{Arrival directions (white dots) of the 70 IceCube events studied in the present work superimposed on the all-sky gamma-ray map, equatorial coordinates. The black star denotes the Galactic center. The color in the map reflects the intensity of the gamma-ray flux with energies above 1~GeV observed by \textit{Fermi}~LAT (\url{https://svs.gsfc.nasa.gov/14090}), with the emission from the Galactic plane clearly seen.
\textit{Fermi} sky map credit: NASA/DOE/Fermi LAT Collaboration.}
\label{f:FermiSky}
\end{figure*}

At the same time, the Galactic contribution has not revealed itself in the distribution of the neutrino arrival directions significantly.
\cite{2016APh....75...60N} searched for the Galactic-disk excess among 19 neutrino events with energies $E>100$~TeV and claimed a $3\sigma$ excess from the range of Galactic latitudes $|b|<10^\circ$. This excess became insignificant with the addition of more data at $E>200$~TeV \citep{ST-Gal}. A detailed analysis of 50 events with $E>60$~TeV \citep{Denton} did not reveal any Galactic excess. Subsequently, IceCube updated \citep{IceCube-HESE-2020,IceCube:muon2021} the arrival directions of events used in these works, shifting some of them by as much as several tens of degrees, cf.\ Figure~3 in \citet{ST-UFN}. Therefore, these early results were deprecated. Subsequent searches for the Galactic-disk excess in the distribution of arrival directions mostly made use of the spectral and directional templates based on simulations of propagation of cosmic rays in the Galaxy and their interactions with the interstellar matter, with the so-called KRA$\gamma$ model \citep{KRAgamma} being one of the most popular. Strict upper limits on the disk component were found for this template by a combined analysis of IceCube and ANTARES data by \citet{IceCubeANTARES-GalPlane}. At the same time, cascade data of IceCube favor a weak, $2\sigma$, excess from the Galactic plane \cite{IceCube:cascades-Gal2sigma}. 
The data sets used in these studies were dominated by low-energy (TeV) events and their results are sensitive to the templates assumed, while cosmic-ray models used to construct these templates contain large uncertainties. 

In the present work, we search for the Galactic-plane enhancement in the distribution of arrival directions of neutrinos with highest energies, $E>200$~TeV, detected and published by IceCube, in a framework that does not rely on any predefined template.

\section{Neutrino Data, Analysis and Result}
\label{sec:analysis}

In \citet{neutradio1}, we constructed a sample of IceCube events with best-fit reconstructed energies above 200~TeV and successfully used it to search for associations between neutrinos and radio blazars. It contained all track-like events in this energy range, information about which had been published by IceCube by that time. We selected the events with reasonably good reconstruction by requiring that the 90\,\% containment area for the reconstructed arrival direction in the celestial sphere $\Omega_{90} < 10$~sq.~deg. By construction, this sample is free of any Galaxy-related directional biases.

For the present study, we choose to use the same sample, supplementing it with newer alert events selected by precisely the same criteria. While the localization condition $\Omega_{90} < 10$~sq.~deg.\ is less important by itself for studies of full-sky anisotropy here than it was for the search of point sources in \citet{neutradio1}, poor reconstruction of the arrival direction is often related to a lower quality of reconstruction of other parameters, including energy. In addition, by keeping the previously defined criteria unchanged we avoid statistical trials and sample tuning for the present analysis, which is always welcome.
The original sample of 56 events used by \citet{neutradio1} is updated using arrival directions and energies published in recent paper by \cite{IceCube:muon2021}. This makes three early events fail our selection criteria, and three new events pass them. The sample is supplemented 
by 14 new events passing the same cuts, reported subsequently by IceCube as GCN\footnote{Gamma-ray Coordinates Network, \url{https://gcn.gsfc.nasa.gov/gcn_main.html}}/AMON\footnote{Astrophysical Multimessenger Observatory Network, \url{https://www.amon.psu.edu}} alerts.
The list of the sample we use, that is of \textit{all} 70 IceCube events which are selected on the basis of the above criteria in 2008--2022, is presented in \autoref{t:icecube_events}.

\begin{table}
\caption{IceCube high-energy neutrino events used in our analysis.
\label{t:icecube_events}
}
\hskip -1.4cm
\footnotesize
\begin{tabular}{ccrrl}
\hline\hline
Date & $E$   & $l$       & $b$       & Reference \\
     & (TeV) & ($\degr$) & ($\degr$) &           \\
(1)  & (2)   & (3)       & (4)       & (5)       \\
\hline
2009-08-13 & $480$ & $155.08$ & $-57.42$ & \cite{IceCube:muon2021} \\
2011-03-04 & \n    & $228.73$ & $6.86$   & \cite{IceCubeEHEAcatalog} \\
2011-07-14 & $253$ & $161.79$ & $-5.4$   & \cite{IceCube-1405.5303} \\
2012-10-11 & $210$ & $326.85$ & $58.17$  & \cite{IceCube-1607.08006} \\
2014-01-22 & $430$ & $304.82$ & $-23.69$ & \cite{IceCube-1510.05223} \\
2016-07-31 & \n    & $343.62$ & $55.55$  & \cite{2017arXiv171001179I} \\
2017-09-22 & $290$ & $195.42$ & $-19.56$ & \cite{IceCubeTXSgamma} \\
2018-09-08 & \n    & $237.2$  & $35.09$  & GCN 23214 \\
\hline
\end{tabular}
\tablecomments{
The set of all 70 IceCube events selected according to criteria, formulated in \autoref{sec:analysis}.
In a few cases, the neutrino energy is not published, but we assume following \citet{neutradio1} that $E>200$~TeV for all events which passed selection criteria for the Extremely High Energy (EHE) IceCube alert sample \citep{IceCubeOldAlerts}.
Shown are eight events only, the complete table is available electronically.}
\end{table}

\begin{figure}
\centering
\includegraphics[width=0.95\linewidth,trim=0cm 0cm 0cm -0.3cm]{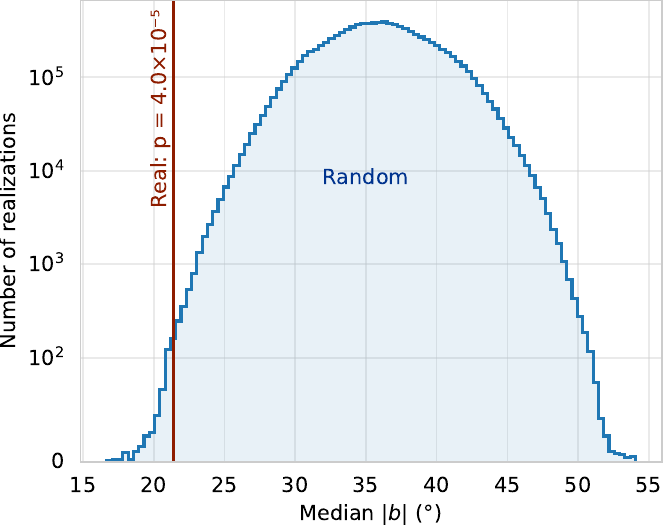}
\caption{Distribution of the median absolute Galactic latitude, $|b|_{\rm med}$, for $10^7$ samples simulated under the assumption of the null hypothesis of the neutrino flux isotropy. The orange line indicates the value of $|b|_{\rm med} = 21^\circ$ for the sample of observed events. Note the log scale in the vertical axis.}
\label{f:median}
\end{figure}

IceCube is located at the South Pole, and its sensitivity to a direction in the sky is determined only by the zenith angle, unambiguously translated to the declination. Therefore, one expects that the arrival directions of reconstructed events are distributed uniformly in right ascension for the isotropic incoming flux. The same is true for the atmospheric background. This assumption is used for modelling of the background in IceCube anisotropy searches \citep[see e.g.][etc.]{IceCube:7yr-sources,IceCubeTXSgamma}. We do the same in what follows in the present paper. For simulations under the null hypothesis, right ascensions of simulated events are generated randomly while declinations are kept unchanged from those in the real data set.

The distribution of arrival directions of the 70 events is shown in \autoref{f:FermiSky}. 
Even by eye, one can note a slight  concentration towards the Galactic Plane.
We turn now to the key point of the paper, quantifying the strength of this effect. To avoid any sensitivity to the assumed shape of the Galactic-plane enhancement, we use the simplest unbinned non-parametric test statistics distinguishing low and high Galactic latitudes, namely the median absolute Galactic latitude of all events, $|b|_{\rm med}$. The potential Galactic-plane excess would correspond to lower $|b|$ than expected for isotropy. For the real data, $|b|_{\rm med}\approx 21^\circ$. In simulations under the null hypothesis, this or a lower value of $|b|_{\rm med}$ occurs very rarely, see \autoref{f:median}.
The (one-tailed) p-value, that is the fraction of simulated sets resulting in this or lower $|b|_{\rm med}$, is $4\times 10^{-5}$. Obtained in a parameter-free way and with the previously fixed full public data set, this p-value does not require trial corrections and determines the final significance of our observations at the level of $4.1\sigma$.

\section{Discussion}
\label{sec:disc}

\subsection{Estimate of the neutrino flux from the Galactic plane}
\label{sec:disc:flux}
\begin{figure}
\centering
\includegraphics[width=0.95\linewidth]{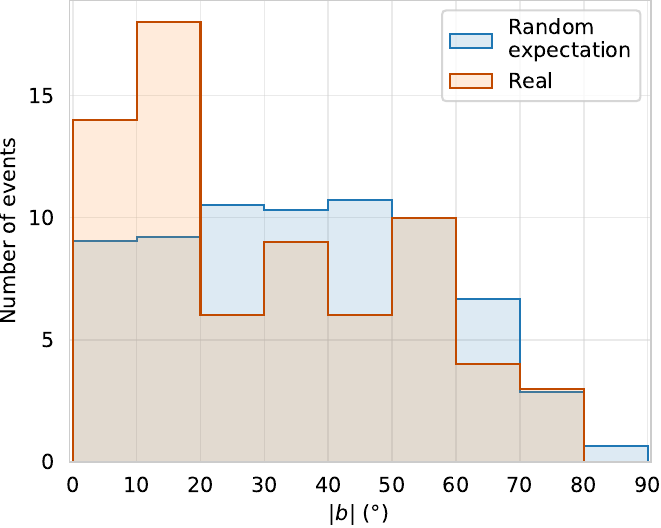}
\caption{Distributions of real (orange) and simulated (blue) events in the absolute Galactic latitude $|b|$ of their arrival directions. The expected number of scrambled events in each bin is estimated by averaging $10^5$ random samples.
}
\label{f:b-distr}
\end{figure}

Once the existence of the excess of high-energy neutrinos from the Galactic plane is established, it is interesting to estimate the corresponding Galactic neutrino flux. To do so, we need to make some assumptions, first of all with respect to the distribution of the excess in the sky. \autoref{f:b-distr} visualizes the distributions in Galactic absolute latitudes $|b|$, comparing real events and  events simulated under the null hypothesis. The Kolmogorov-Smirnov (KS) probability that the observed and simulated events are drawn from the same underlying Galactic latitude distribution is $p_{\rm KS}\approx 7\times 10^{-4}$. KS test detects a wide range of possible differences in distributions, and is by necessity less sensitive to changes of the average value.

Motivated by \autoref{f:b-distr}, we assume in our further estimates that the excess comes from the band $|b|\le 20^\circ$, where 32 of 70 events are observed while 18 are expected for an isotropic distribution. The 68\,\% binomial confidence interval for the excess is $13\pm4$ events. We stress that this assumption is used for interpretation only and does not affect the main result of the paper presented in \autoref{sec:analysis} and its statistical significance. A more detailed but model-dependent study should include modelling of the distribution by a sum of the assumed Galactic, extragalactic and background components with fitted coefficients, like it was done by \citet{ST-Gal} with early data. 

Even at these high energies, a large number of events in the sample are not astrophysical since background atmospheric neutrinos and muons still pass the strict selection criteria. \citet{neutradio2} estimated that about 1/3 of the events in this sample are atmospheric. This estimate was based on simulations reported for particular classes of contributing events. Since the number of events of each class in the final sample is small, this fraction may fluctuate considerably. Within these expected fluctuations, the fraction 2/3 of astrophysical neutrinos in the present sample can be used as well. Taking it into account, we estimate that about $(28\pm 9)$\% of astrophysical events above 200~TeV are Galactic and the corresponding Galactic-disk neutrino flux at $E=200$~TeV is 
$\sim 2.3 \times 10^{-16}$~TeV$^{-1}$cm$^{-2}$s$^{-1}$sr$^{-1}$. Here, we used the most recent power-law spectral fit obtained by IceCube from muon-track analysis \citep{IceCube:muon2021} and took into account the solid angle spanned by the $\pm 20^\circ$ band around the Galactic equator. 

\subsection{Comparison with the 10-year IceCube public catalog}
\label{sec:disc:10yr}

We found the Galactic excess by analyzing the $E>200$~TeV data set, but one expects that the Galactic contribution exists also at lower energies. At which energies it is important? This question is hard to answer precisely because of the large atmospheric background progressively dominating the event samples at lower energies.
In addition, neutrino energies have not been published for most IceCube events. Therefore, we do not attempt to assign statistical significance to the results obtained in the present subsection.

To perform a rough estimate, we used the public 10-year  IceCube catalog of all track-like events \citep{IceCube10yrData,IceCube10yrDataPaper} for this auxiliary analysis. Note that in this catalog, only reconstructed muon energies $E_\mu$ are given, which are considerably lower than the neutrino energies $E$. The relation between $E_\mu$ and $E$ differs between events and is not readily available. We select the ``northern-sky'' events, 
$\mathrm{Dec}>-5^\circ$ \citep[see e.g.][]{IceCube:7yr-sources}, to suppress the background of atmospheric muons, and select events with $\Omega_{90} < 10$~sq.~deg.\ for the reasons discussed in \autoref{sec:analysis}.

\begin{figure}
\centering
\includegraphics[width=\linewidth]{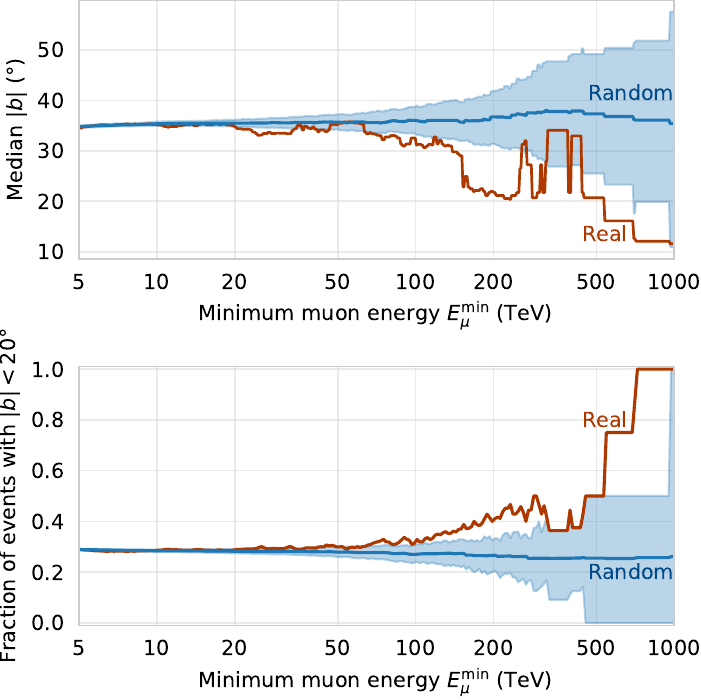}
\caption{The median absolute Galactic latitude of events with muon energy above $E_\mu^\mathrm{min}$ in the 10-year IceCube muon-track catalog (top), and their fraction within $20^\circ$ from the Galactic plane (bottom).
Simulated confidence intervals are shown at a 68\,\% level. The apparent wobble at high energies is due to a progressively smaller number of events.
}
\label{f:IceCube10yr}
\end{figure}

As shown in \autoref{f:IceCube10yr} for the median galactic latitude, higher-energy events become closer to the galactic plane, aside from random fluctuations.
\autoref{f:IceCube10yr} also presents
the fraction of events close to the Galactic plane in the data and in similar data sets simulated under the null hypothesis. To define the proximity to the Galactic plane we use the same $|b|<20^\circ$ cut as for all our estimates in this discussion section. One can see that the Galactic excess starts to be seen from muon energies of a few tens of TeV, just where the astrophysical contribution starts to be noticeable above the atmospheric background \citep{IceCube:muon2021}. The excess becomes the most pronounced in the hundreds of TeV, reaching up to 20\,\% of all events. According to \cite{IceCube:muon2021}, astrophysical events constitute 50\%\,--\,80\% of detections in this energy range, and the excess is consistent with the main $(28\pm 9)$\% estimate from \autoref{sec:disc:flux}.
It is interesting to note that two highest-energy events both have $|b|<20^\circ$, though it may agree with expected fluctuations.

\subsection{Origin of the Galactic neutrinos}
\label{sec:disc:origin}

The excess in the neutrino arrival directions is wider than the expected thickness of the stellar disk and therefore suggests that at least a part of the Galactic high-energy neutrinos originate in cosmic-ray interactions with the diffuse matter and radiation. This contribution is indeed guaranteed \citep[e.g.][]{Neronov:interstellar,Neronov:Milky} by the undoubted existence of both the energetic cosmic rays and the interstellar matter, but it was expected to give only a moderate fraction of the total flux measured by IceCube. At the same time, individual cosmic-ray accelerators, ``PeVatrons'' \citep[cf.\ e.g.][and references therein]{Kheirandish:2020upj}, are expected to contribute to the Galactic neutrino flux as well. Some of them have been possibly observed in gamma rays above 100~TeV 
\citep[see e.g.][]{Tibet-Crab-100TeV,HAWC56-100,LHAASO-12sources}. 
Of particular interest is the 
observation by \citet{Carpet-Cocoon} of a gamma-ray flare above 300~TeV from the Cygnus region, coinciding in direction and time with a 150-TeV neutrino detected by IceCube and possibly associated with a gamma-ray binary system \citep{BykovCygnusPSR}.

Recent IceCube and ANTARES searches for the neutrino excess from the Galactic plane \citep{IceCubeANTARES-GalPlane,IceCube:cascades-Gal2sigma} were based on the KRA$\gamma$ template \citep{KRAgamma} for the diffuse neutrino flux from cosmic-ray interactions and assumed the cosmic-ray spectral cutoff is at either 5~PeV or 50~PeV. The obtained upper limits, which depend on the assumed spectrum and cutoff, are saturated by much lower energies than we consider here. Cosmic-ray measurements \citep[e.g.\ by TALE,][]{TALE1,
TALE2} indicate the continuation of the Galactic cosmic-ray proton spectrum to higher energies than assumed in the construction of these templates. Note that our main result, \autoref{sec:analysis}, presents direct evidence of the Galactic-plane excess which does not make use of any template. 

The excess of neutrinos produced by the ``sea'' of Galactic cosmic rays interacting with gas is expected to be more concentrated towards the Galactic plane than the one we observe, see e.g. \citet{astro-ph/0701856}. At the same time, cosmic rays close to relatively nearby young sources could give a contribution at larger Galactic latitudes, similar to the observed one.

\subsection{Diffuse Galactic neutrinos and diffuse Galactic gamma rays}
\label{sec:disc:gamma}

High-energy neutrino production is accompanied by emission of gamma rays in the same energy band as neutrinos \citep[see e.g.][and references therein]{ST-UFN}. For the most common hadronic-interaction mechanisms, where neutrinos are born in decays of charged $\pi$ mesons, gamma rays come from decays of their neutral counterparts. It is remarkable that the Galactic-plane diffuse gamma-ray flux above 100~TeV has been recently observed \citep[Tibet-AS$\gamma$ experiment,][]{Tibet-GalDiffuse}. If these gamma rays have hadronic origin, then the corresponding Galactic neutrino flux is expected
\citep[e.g][]{Tibet-neutrino1,
Tibet-neutrino2,Tibet-neutrino3,
Tibet-neutrino4,Tibet-neutrino5}. Simple relations for kinematics of $\pi$-meson decays, collected e.g.\ in Sec.~3.1 of \citet{ST-UFN}, allow one to relate the fluxes of neutrinos, $dN_\nu/dE$, at the neutrino energy $E$, and of gamma rays, $dN_\gamma/dE_\gamma$, at the gamma-ray energy $E_\gamma$, as
\[
E_{\gamma}^{2} \frac{dN_{\gamma }}{dE_{\gamma }} 
\approx
A E^{2} 
\left. \frac{dN_{\nu }}{dE} \right|_{E=E_{\gamma}/2} 
,
\]
with $A\approx 2/3$ ($4/3$) for the $pp$ ($p\gamma$) interactions, respectively. The neutrino flux at $E>200$~TeV we discuss here thus corresponds to the flux of gamma rays with energies $E_\gamma>400$~TeV. Assuming $pp$ interactions and the neutrino flux estimated in \autoref{sec:disc:flux}, the expected Galactic-plane flux of photons above 400~TeV is $\sim(7.8 \pm 3.3)\times 10^{-17}$~TeV$^{-1}$cm$^{-2}$s$^{-1}$sr$^{-1}$. This estimate fits well with the measurement of the Galactic-plane gamma-ray flux at 398~TeV$<E_\gamma<$1000~TeV reported by \citet{Tibet-GalDiffuse}. A more detailed comparison would require particular quantitative assumptions about the distribution of sources in the Galaxy and of the shape of the emitted spectrum. This information is needed to account for the absorption of gamma rays due to pair production on the cosmic background photons, \citep{Nikishov1962}, 
important for energies above 100~TeV. \citep{AhlersMurase-gamma-Gal,Winter-gamma-Gal,KalashevST-gamma-Gal}.
We note that the Tibet gamma-ray signal matches the neutrino signal also in the spatial extension, $\sim \pm 10^\circ$ around the Galactic plane.

\subsection{Galactic and extragalactic contributions}
\label{sec:disc:gal-egal}

The Galactic disk component we find here is an important but not unique contribution to the astrophysical high-energy neutrino flux. It coexists with a well-established contribution from extragalactic sources described in \autoref{sec:intro}. There could exist also other Galactic contributions which do not produce the disk-related anisotropy, e.g.\ if the neutrino-producing regions are located either in the immediate neighbourhood of the Solar system \citep{LocalBubble1,LocalBubble2} or in the extended halo of the Galaxy \citep[e.g.][and references therein]{halo1,no-halo1,halo2,
no-halo2}; they are not expected to reveal themselves in the present study. 

In general, the disk component should be taken into account in the modelling of the background for studies of these other contributions. However, we emphasise that when (and only when) an isotropic complete catalog of sources is used, the anisotropy of the background does not affect the resulting significance. Examples include a complete sample of VLBI-selected extragalactic radio sources used to study the blazar-neutrino connection \citep{neutradio1,neutradio2} or blazar-gamma-ray cross-identifications \citep{Kovalev09}.

\section{Summary}
\label{sec:summary}

We found $4.1\sigma$ evidence (p-value of $4\times 10^{-5}$) for the existence of an anisotropic component of the neutrino flux above 200~TeV, coming from low Galactic latitudes. It constitutes about one third of the total astrophysical flux at these energies. The median Galactic latitude value of the events, $|b_{\rm med}|\approx 21^\circ$, reflects the superposition of neutrinos from the Galactic disk, nearby sources, extragalactic, and atmospheric contributions. The estimated Galactic diffuse neutrino flux agrees with multimessenger expectations for the Galactic diffuse gamma-ray flux above 400~TeV found recently by Tibet-AS$\gamma$ \citep{Tibet-GalDiffuse}. 
Future studies with huge neutrino telescopes
are needed to unambiguously pinpoint the origin of the Galactic-disk high-energy neutrinos found in the present work.
At least two components, Galactic (this study) and extragalactic \citep[from blazars, e.g.][]{neutradio1,neutradio2}, are found to significantly contribute to the observed neutrino flux.
The neutrino sky may well be even more complex, like the electromagnetic one.

\begin{acknowledgments}
We thank Alan Roy for useful comments on the manuscript.
This work is supported by the Ministry of science and higher education of Russia under the contract 075-15-2020-778.
\end{acknowledgments}

\bibliographystyle{aasjournal}
\bibliography{neutgalaxy}
\end{document}